\documentclass[12pt]{article}

\usepackage{graphicx}
\usepackage{epsfig}
\usepackage{bm}

\newcommand{\mb}[1]{\mbox{{#1}}}

\newcommand{\bchi}{\bar{\chi}}
\newcommand{\f}[2]{\frac{#1}{#2}}

\newcommand{\delm}{\partial_\mu}
\newcommand{\del}{\partial}

\newcommand{\kleiner}{\scriptsize}

\newcommand{\D}{\displaystyle}

\newcommand\bc{\begin{center}}
\newcommand\ec{\end{center}}
\newcommand\bt{\bc\begin{tabular}}
\newcommand\et{\end{tabular}\ec}

\newcommand\ban{\begin{eqnarray}\nonumber}

\newcommand\baa{\begin{array}}
\newcommand\eaa{\end{array}}
\newcommand\be{\begin{equation}}
\newcommand\ee{\end{equation}}
\newcommand\ba{\begin{eqnarray}}
\newcommand\ea{\end{eqnarray}}

\newcommand\la{\langle}
\newcommand\ra{\rangle}

\newcommand\bmm{\begin{minipage}}
\newcommand\emm{\end{minipage}}

\def\IJMP#1{Int. J. Mod. Phys.~{\bf #1}}

\def\NP#1{Nucl. Phys.~{\bf #1}}

\def\PL#1{Phys. Lett.~{\bf #1}}
\def\PR#1{Phys. Rev.~{\bf #1}}
\def\PRP#1{Phys. Rep.~{\bf #1}}

\def\PTP#1{Prog. Theor. Phys.~{\bf #1}}
\def\PTPS#1{Prog. Theor. Phys. Suppl.~{\bf #1}}

\def\ZP#1{Z. Phys.~{\bf #1}}

\def\mpla#1{Mod. Phys. Lett. {\bf A#1}}

\def\plb#1{Phys. Lett. {\bf #1B}}

\begin{document}

\title{Finite-Temperature Gluon Condensate with \\Renormalization Group Flow
  Equations}

\author{B.-J. Schaefer, O. Bohr  and J. Wambach\\[2mm]
 {\small \em {Institut f\"ur Kernphysik, TU Darmstadt, 
        D-64289 Darmstadt, Germany}}
}

\maketitle
\begin{abstract}
  \  Within a self-consistent proper-time Renormalization Group (RG)
  approach we investigate an effective QCD trace anomaly realization
  with dilatons and determine the finite-temperature behavior of the
  gluon condensate.  Fixing the effective model at vanishing
  temperature to the glueball mass and the bag constant a possible
  gluonic phase transition is explored in detail. Within the RG
  framework the full non-truncated dilaton potential analysis is
  compared with a truncated potential version.
\end{abstract}
\noindent{PACS: 05.10.Cc, 64.60.A, 14.70.D}

\section{Introduction}

Perturbative QCD studies and lattice simulations demonstrate that 
the fundamental quark and gluon degrees of freedom are the relevant ones 
at high temperatures. Since these degrees of freedom are confined in the 
low-tem\-pera\-ture regime there must be a quark and/or gluon deconfinement
phase transition at high temperature.

Difficulties arise in the description of this phase transition due to
the absence of local parameters which can be linked to confinement.
QCD at low energies can be approxi\-mated by effective models which have
to satisfy the observed symmetry properties and anomaly structure of
the theory. Besides the important approxi\-mate and spontaneously broken
chiral symmetry at low energy, the classical QCD-Lagrangian exhibits
an additional dilatation symmetry or scale invariance in the limit of
vanishing current quark masses which is embedded in a larger conformal
group.  This symmetry is broken at the quantum level by radiative
corrections -- the so-called trace anomaly.

In order to mimic the trace anomaly in an effective framework one
introduces a scalar ($J^{PC} = 0^{++}$) dilaton field $\chi$ with
scaling dimension one and a logarithmic potential with coupling $h$
of the form~\cite{sche,grom}

\ba\label{potchi} 
V &=& h \left(\frac{\chi}{\chi_0}\right)^4 \ln
\left(\frac{\chi}{\chi_0 e^{1/4}}\right)\ . 
\ea 

This potential breaks scale invariance and leads therefore to a finite
vacuum expectation value (VEV) $\chi_0 \equiv \la 0|\chi |0\ra$. In
this way the anomalous breaking of scale invariance is effectively
taken into account.  This can be seen by identifying the trace of the
energy-momentum tensor, $\theta_\mu^{\mu\ \mbox{\kleiner eff}}$, of
the effective theory with the trace of the modified (symmetric)
energy-momentum tensor $\theta_\mu^{\mu\ \mbox{\kleiner QCD}}$ of
massless QCD. The trace is related to the divergence of the dilatation
current $J_\mu^{\mbox{\kleiner dil}}$, which vanishes at tree level
where fluctuations are omitted:

\ba\label{trace} \partial^\mu J_\mu^{\mbox{\kleiner dil}} =
-\frac{h}{\chi_0^4}\chi^4 = \theta_\mu^{\mu\ \mbox{\kleiner eff}}
&\stackrel{!}{=}& \theta_\mu^{\mu\ \mbox{\kleiner QCD}} = \f
{\beta(g)^{\mb{\kleiner QCD}}}{2g} G_{\mu \nu} G^{\mu \nu}\ .
\ea

\noindent
Here $G_{\mu \nu}$ denotes the non-Abelian field strength tensor,
$\beta(g)^{\mb{\kleiner QCD}}$ the QCD beta (Gell-Mann--Low) function
and $g$ is the gauge coupling.

Beyond tree level, however, massless QCD is no longer scale-invariant.
Due to the renormalization, a new dimensionful cut-off hidden in the
QCD beta function must be introduced. Despite this fact, the anomaly
does not depend on the chosen regularization scheme. The effective
realization of the trace anomaly can be achieved by equating the
divergence of the dilatation current with the dilaton field. This
yields a differential equation for the $a$ $priori$ unknown effective
dilaton potential whose solution is
Eq.~(\ref{potchi}).

This effective realization allows for an identification of the gluon
condensate $\la 0 | G_{\mu\nu} G^{\mu\nu} |0 \ra$ with the VEV
$\chi_0$ of the effective theory with broken scale invariance via
Eq.~(\ref{trace}).  Following the suggestion of Campbell et
al.~\cite{camp} $\chi_0$ can be regarded as an effective order
parameter for the deconfining phase transition.  The authors find a
strong correlation with the chiral phase transition, arguing that a
first-order gluonic phase transition might drive the usual
second-order chiral phase transition for two flavors, thereby changing
it to first order.  Their conclusion is that the quark condensation
temperature, $T_q$, for the chiral phase transition is always smaller
or equal to the gluon condensation or confinement temperature $T_g$.
Even for large $N_c$ and fixed number of flavours the quark transition
would be driven by the gluon transition, because $T_q$ and $T_g$
coincide in this case with the critical temperature of a first-order
gluon transition.

In order to investigate these findings we perform a finite-temperature
analysis of the effective theory within the framework of a proper-time
renormalization group (PTRG) approach. With the logarithmic
interaction potential in Eq.~(\ref{potchi}) the Lagrangian becomes
non-renormalizable. One possible regularization of the infinite $T=0$
contributions would be the introduction of a cut-off parameter
$\Lambda$ which determines the scale up to which the effective
description of the theory is valid, similar to a model
parameterization of the Nambu--Jona-Lasinio type.  Such a new
dimensionful cut-off would violate the scaling properties of the
effective theory. In order to maintain the desired scaling behavior,
one has to give the cut-off-parameter a conformal weight of unit one
by some mechanism.

For the purely gluonic Lagrangian specified by the potential in
Eq.~(\ref{potchi}), a direct comparison of the free energy or
effective potential at the origin, and at the local minimum, needed in
a perturbative approach, is not possible due to the breakdown of the
Taylor expansion of the potential itself.  It is therefore impossible
to determine perturbatively the order (first or second) of the phase
transition signalled by the disappearance of the scalar dilaton field
$\chi_0$ as the order parameter.

In order to address the question of the order of the gluon phase
transition, Campbell et al.~\cite{camp} introduced an additional term
proportional to $\chi^n$ ($n<4$) to the effective potential in
Eq.~(\ref{potchi}) with an $a$ $priori$ unknown temperature-dependent
coefficient.  They chose phenomenologically a monotonically increasing
function of temperature for this coefficient. Neglecting effects of
order ${\cal O} (T^4 )$ in the finite-temperature contributions to the
effective potential, a first-order transition was found, as one would
expect from lattice simulations \cite{kars}.

Within the PTRG approach we can circumvent these difficulties by
directly calculating the full non-truncated effective dilaton
potential for any value of the field and not just at the minimum,
$\chi_0$. This in principle allows to analyze the order of the phase
transition.

The outline of the paper is as follows: In Sec.~\ref{sect2} we
introduce a self-consistent RG method combined with a Schwinger
(proper-time) regularization. We derive flow equations for two
different scenarios. The dilaton model is discussed without any
polynomial potential truncation in Sec.~\ref{pureg}. This full
potential calculation is compared to an analysis in a truncated
potential approxi\-mation, followed by an error estimation in
Sec.~\ref{truncg}.  Finally, in Sec.~\ref{concl} we present our
conclusions.


\section{The Renormalization Group approach} 
\label{sect2} 

The RG approach used in the present work is based on a perturbative
one-loop expression for the effective potential, which is regularized
by Schwinger's proper-time representation of the divergent logarithm.
The usual Schwinger proper-time integral is modified by a regulator or
blocking function $f_k$ in the integrand, thus rendering the resulting
flow equation infrared (IR) and ultraviolet (UV) finite.  In this way
the IR flow scale $k$ is introduced~\cite{flor,liao}. The flow
equation\footnote{For an introductory review see e.g.~\cite{bagn},
  Chap.~3 and 4.2 .} that describes the scale dependence of the
action $\Gamma_k$ reads\footnote{In general the remaining trace on the
  $rhs$ of this flow equation runs over all inner degrees of freedom.
  For only one scalar degree of freedom we can omit the trace. } \ba
\label{master}
k \frac{\del \Gamma_k}{\del k} &=& -\frac{1}{2}\int\limits_0^\infty
\frac{d\tau}{\tau} \left(k\frac{\del f_k}{\del k}\right) \mbox{Tr}\;
e^{\D -\tau \Gamma^{(2)}_k} 
\ea 
\noindent
where on the $rhs$ a renormalization group improvement is performed by
a replacement of the bare second derivative of the action w.r.t. the
fields with the corresponding running expression $\Gamma^{(2)}_k$
\cite{bohr}.  This replacement improves the one-loop flow equation and
corresponds to a resummation of higher Feynman graphs i.e. daisy and
superdaisy diagrams~\cite{liao,dola}, similar to a Schwinger-Dyson
resummation. In a plane-wave basis the blocking function $f_k$ serves
as a momentum regulator similar to the regulator $R_k$ in the 'Exact
Renormalization Group' (ERG)~\cite{wett} and selects only modes which
are peaked around the scale $k$~\cite{mazz}. The direct connection to
the cut-off $R_k$ in the ERG formalism is still missing but the issue
whether this proper-time flow converges to the full effective action
will not be discussed here. We also omit a detailed analysis of the
scheme dependence introduced by the choice of the regulator function
\cite{mazz,bona}.

The regulator has to fulfil basically two conditions: The modified
action $\Gamma_k$ should tend to the full action in the IR limit,
requiring that $f_{k\to 0}(\tau \to \infty) \to 1$. In this way the
cut-off is removed and all quantum fluctuations are taken into
account. In addition we have to set $f_k (\tau =0)=1$ for arbitrary
$k$. This does not regularize the UV regime ($\tau =0$) which is not
necessary anyway if we start the evolution at a finite (large) UV
scale $\Lambda$.  Finally, for the derivation of all flow equations we
will employ the following choice for the blocking function $f_k$ in
$d$ space-time dimensions: \ba k\frac{\del f_k}{\del k} &=&
-\frac{4}{d \Gamma(d/2)} (\tau k^2)^{d/2+1} e^{-\tau k^2} \ea
corresponding to the notation $f_k^{(1)}$ in \cite{bohr}.

Of course, it is not possible to solve the full RG equation
(\ref{master}) without truncation. In order to obtain a tractable set
of coupled flow equations we perform a derivative expansion of the
effective action which takes the generic form

\ba 
\Gamma_k [\phi]
&=& \int d^dx \left\{ V_k(\phi) + \f 1 2 Z_k (\phi) (\delm \phi)^2 +
  \ldots \right\}\ . 
\ea 

In this work we consider the local potential approximation (LPA) in
which the running of the wavefunction renormalization $Z_k$ is
neglected ($Z_k$ is set to unity) and only the effective potential
$V_k$ is taken into account.

The task is to solve the resulting coupled non-linear flow equation
for the potential $V_k$ by starting in the ultraviolet with $a$
$priori$ unknown initial conditions and integrate towards zero
momentum with respect to the scale $k$. In this way all quantum
fluctuations are effectively taken into account. The initial
conditions at the UV scale are determined in such a way that
calculated physical quantities in the IR are matched to predetermined
values.

The generalization of our RG method to finite temperature within the
Matsubara formalism is straightforward. Details can be found in
Refs.~\cite{bohr,papp,scha}. In the zero-momentum components of the
loop integration of Eq.~(\ref{master}) one has to introduce bosonic
Matsubara frequencies $\omega_n = 2\pi n T$. In the LPA the resulting
momentum integrals can be performed analytically, leading to
fractional powers in the finite-temperature threshold functions and
hence to the dimensional reduction phenomenon~\cite{scha,tetr}.

\section{The full dilaton potential}
\label{pureg} 

The aim of the next two sections is to derive and solve numerically
flow equations for the gluon condensate from an effective dilaton
Lagrangian with a logarithmic interaction potential. In this section
we consider the full dilaton interaction potential without any
polynomial potential truncation and will compare the results with a
truncated potential calculation in the next section.

As motivated in the introduction we start with the following effective
Lagrangian in Euclidean space\footnote{We neglect in this work any
  quark i.e.~mesonic degrees of freedom.} 

\ba\label{pureglueL} {\cal L} &= &
\frac{1}{2} ( \delm \chi )^2 + V_0 (\chi )\qquad\mbox{with}\qquad V_0
(\chi ) = h \left(\frac{\chi}{\chi_0} \right)^4\ln \left(
  \frac{\chi}{\chi_0 e^{1/4}}\right)\ .
\ea 

The potential $V_0$ is parameterized by two constants. The depth is
determined by the coupling $h$ and the curvature is given by the
minimum $\chi_0$.  The minimum is asymmetric and small oscillations
about $\chi_0$ can be interpreted as a massive scalar gluonium
(glueball) state.  Difficulties arise around $\chi =0$ since the
potential is unstable (actually not defined as mentioned in the
introduction).  Therefore, a particle interpretation at the origin is
also no longer possible.  For Campbell et al.~\cite{camp} this
phenomenon is an indication that glueballs are no longer the relevant
degrees of freedom at that point.  All physical hadrons (if there is
any phase transition in the effective model) are dissolved into their
perturbative gluon content and this phase corresponds to a deconfined
phase.

At the minimum $\chi_0$ the potential is equal to $V_0 (\chi_0)
= - h/4$. This allows for an identification of the coupling $h$ with
the bag constant $B$ which is the negative of the energy density of
the vacuum $\epsilon_{vac}$ via the relation

\ba\label{bagkonst} 
B = V_0 (0) - V_0 (\chi_0) &=& -\epsilon_{vac} = \frac{h}{4}\ .  
\ea

Since we are able to calculate the potential at the origin, this
relation can be used to determine $B$ in the IR.

The first derivative of the potential w.r.t.~the scalar field $\chi$,
denoted in the following with a prime, $V_0' \equiv dV_0 / d\chi$,
vanishes when evaluated at the minimum. The second derivative
$V_0''(\chi_0 ) = 4h/\chi_0^2 = m^2_\chi$ yields the tree-level
glueball squared mass, $m^2_\chi$.  In this work the glueball mass is
taken as an input.  We choose values around $m_\chi = 1.5$ GeV which
is motivated by recent lattice results~\cite{bita}.  The experimental
situation is not clear up to now but glueball candidates in the mass
region of $1.5 \ldots 1.8$ GeV are favored~\cite{part}.  For instance,
a possible candidate could be the $f_0 (1500)$ meson\footnote{Other
  possible candidates are the $f_0 (1370)$ and $f_0(1710)$ mesons.
  Recently, the possibility of mixing of these mesons with nearby $q
  \bar{q}$ states has also been discussed.}.  The value of the bag
constant $B$ which serves as a second input is not well known but
ranges between $B^{1/4} = 0.14 \ldots 0.24$ GeV.  Thus, using e.g.
$B^{1/4} = 0.24$ GeV and $m_\chi^2 = 1.5^2$ GeV$^2$ at the UV scale
(tree level) yields $h^{1/4} = 0.34$ GeV and $\chi_0 = 0.154$ GeV. In
Ref.~\cite{soll} it is observed that the temperature scale, where
thermal excitations become important, is determined by the value of
the bag constant at $T=0$. For $B^{1/4} = 0.14$ GeV, the onset of a
significant shift of the minimum is seen at $T \sim 0.25$ GeV while
for $B^{1/4} = 0.24$ GeV this happens only for temperatures above
$T\sim 0.4$ GeV.  In fact, it was found in \cite{soll} that the value
for the ``critical'' temperature $T_c$ is dominated by the value of
the bag constant. A larger bag constant results in a higher critical
temperature, although the authors cannot determine a real first- or
second-order phase transition.  From the difference between the RG
evolved potential in the IR at the true minimum $\chi_0$ and the
potential at the origin we can calculate the bag constant by means of
Eq.~(\ref{bagkonst}) and find a similar dependence of the ``critical''
temperature on the bag constant.

In order to investigate the behavior of the dilatons in a RG picture
we apply the method presented in Sec.~\ref{sect2} to the Lagrangian of
Eq.~(\ref{pureglueL}). For the dilaton potential in the LPA with $d$
Euclidean space-time dimensions and the smearing function\footnote{See
  also for further details and definitions Refs.~\cite{papp,bohr}.}
$f^{(1)}_k$ we obtain the following flow equation:
 
\ba\label{flowg0} 
\partial_t V_k &=& \f{2}{(4\pi)^{d/2} \Gamma (d/2 )} \f{k^d}{d}
\f{1}{(1 + \tilde{V_k}'')}\ , 
\ea

\noindent
where $t = \ln (k/\Lambda)$ and the tilde indicates the rescaled
quantity w.r.t.~the IR squared scale $k^2$, e.g. $\tilde{V}''_k =
V''_k / k^2$.  In order to solve this coupled equation we discretize
the field $\chi$ for a general potential $V(\chi)$ on a grid. To close
the system of equations we need to derive an equation for the first
derivative $V'_k$~\cite{adam}. This generates higher derivatives of
the potential on the $rhs$ of Eq.~(\ref{flowg0}) up to $V'''_k$.
These higher derivatives are then numerically determined by matching
conditions, based on a Taylor-expansion of $V_k$ and $V'_k$ at
intermediate grid points.  In this way we obtain a highly coupled
closed system of flow equations which can be solved numerically with a
5th order Cash-Karp Runge-Kutta method~\cite{pres}.  The advantage of
closing the system in this way is that we do not need higher than
third-order derivatives. Since the expansion of a logarithmic
potential at the origin in a power series leads to singular
coefficients for powers larger than three this signals the breakdown
of a perturbative treatment of the model, as already mentioned in the
introduction. Via the application of the non-perturbative RG method we
can circumvent these singularities.

The initial conditions for the coupling $h$ and minimum $\chi_0$ are
determined at the UV scale $\Lambda$ by fits to the glueball mass
$m^2_\chi$ and the bag constant $B$ at the end of the evolution in the
IR.  We have chosen $\Lambda = 2.0$ GeV, $\chi_0=0.137$ GeV and a
coupling $h = (0.51 \mbox{GeV})^4$ and start with a logarithmic dilaton
potential (tree level). These initial values result after the
$k$-evolution in a bag value of $B^{1/4} = 0.276$ GeV and a glueball
ball mass of $m_\chi = 1.502$ GeV.

\begin{figure}[!ht] 
  \centerline{\hbox{ \psfig{file=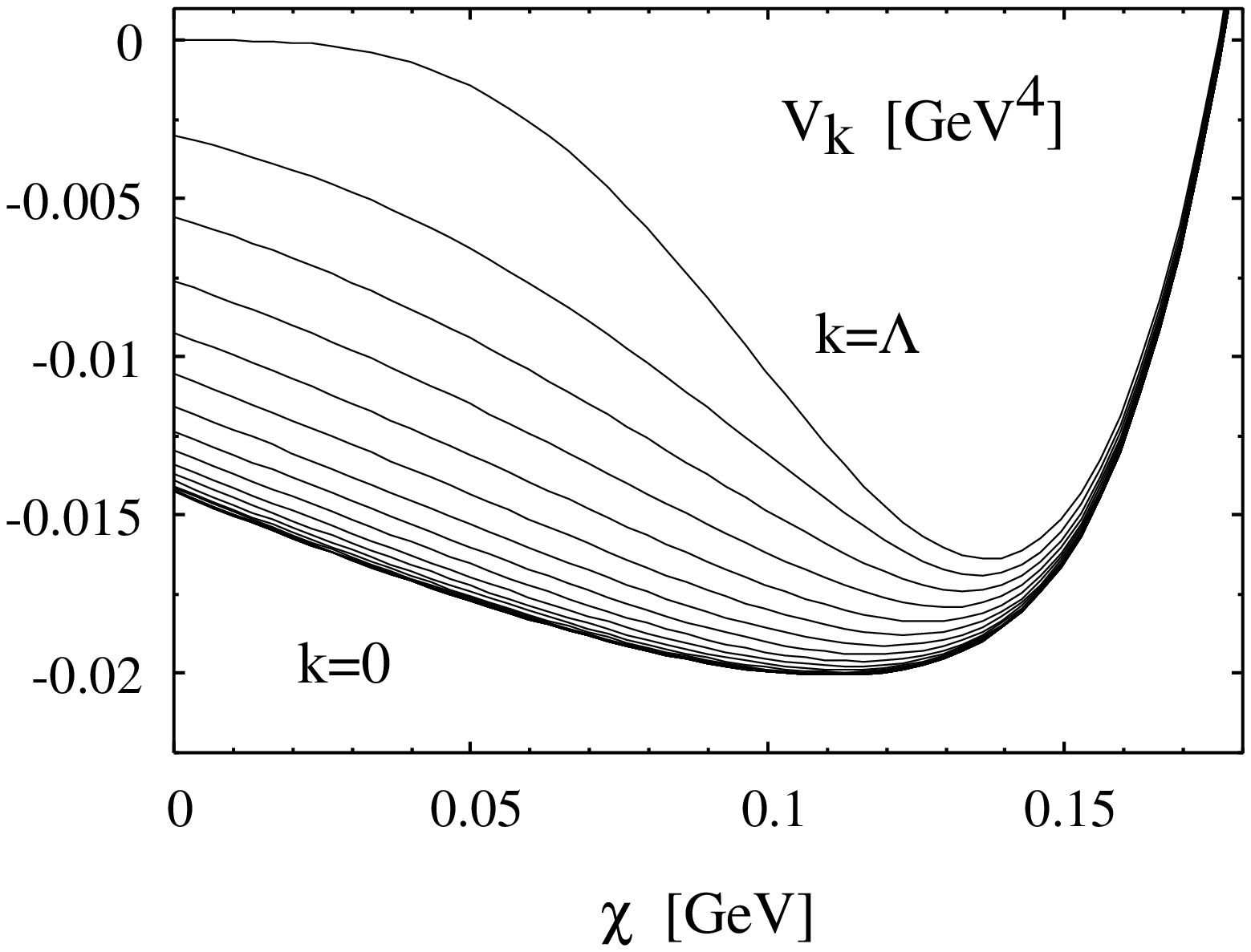, width=7cm,angle=0}
      }\hfill \psfig{file=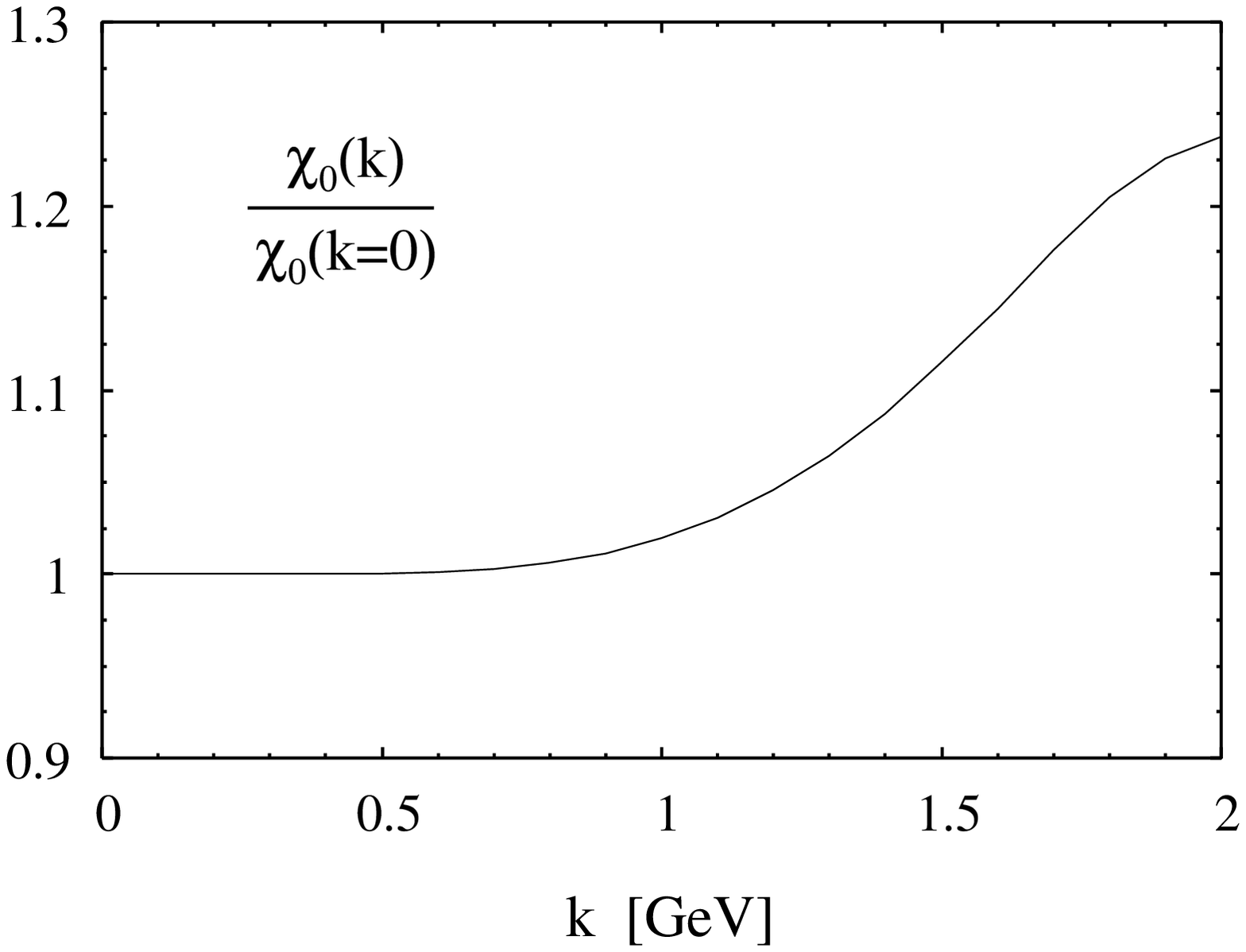,width=7cm,angle=0} }
\caption{\label{figkfullpot}  
  The $k$-evolution of the potential $V_k$ (left panel) versus $\chi$
  and the $k$-evolution of the normalized minimum
  $\chi_0(k)/\chi_0(k=0)$ versus $k$ (right panel).  (h$^{1/4}$ = 0.51
  GeV, $\Lambda = 2.0$ GeV , $\chi_0 (\Lambda) =0.137$ GeV). }
\end{figure} 

Due to the radiative corrections (quantum fluctuations) the shape of
the effective logarithmic potential is altered when evolving towards
the IR as shown in Fig.~\ref{figkfullpot} (left panel).  In the IR,
the potential becomes convex for all $\chi$ values. This is consistent
with the definition of the scale anomaly.

The structure of the threshold functions in the flow
Eqs.~(\ref{flowg0}) is very similar to those for the $O(N)$ symmetric
model~\cite{bohr}. Due to a negative curvature of the potential for
small $\chi$ which is typical for broken symmetries, poles can occur
in the threshold functions. For instance, $V''_k/k^2$ starts with
negative values for small field amplitudes $\chi$ during the
$k$-evolution.  That is the reason why there is a correlation of the
initial values $h$, $\chi_0$ and the UV scale $\Lambda$ and why they
cannot be chosen independently. To avoid a pole in the threshold
function we have to choose initial values in such a way that
$V''_k/\Lambda^2$ is larger than -1 and $V''_k$ becomes positive for
all field amplitudes $\chi$ during the evolution towards the IR.  In
the right panel of Fig.~\ref{figkfullpot} the $k$-evolution of the
normalized minimum is presented. One recognizes that the evolution
almost stops around $k \approx 0.4$ GeV.

\begin{figure}[!ht] 
  \centerline{\hbox{ 
       \psfig{file=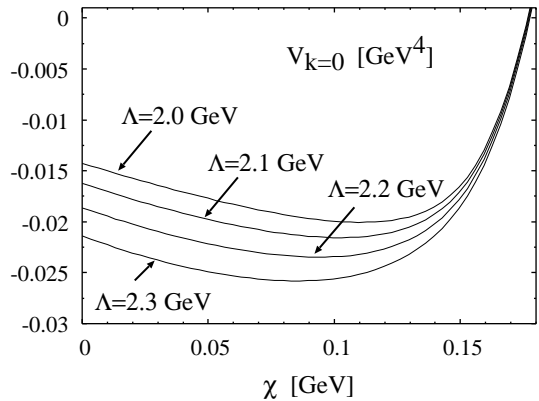,width=7cm,angle=0} }}
\caption{\label{figcutoff} 
  The evolved effective potential $V_{k=0}$ versus $\chi$ for various
  values of the UV cutoff $\Lambda$ (h$^{1/4}$ = 0.51 GeV, $\chi_0
  (\Lambda) =0.137$ GeV). }
\end{figure}

In order to estimate the theoretical uncertainty of these results we
show in Fig.~\ref{figcutoff} the UV cutoff dependence of the evolved
effective potential $V_{k=0}$ for values $\Lambda = 2.0 - 2.3$ GeV at
fixed coupling $h^{1/4} = 0.51$ GeV and a fixed minimum $\chi_0 =
0.137$ GeV.
\begin{figure}
\begin{center}
\begin{tabular}{lll}
\hline
\footnotesize$\qquad\Lambda \quad\mb{[GeV]}$ &\footnotesize$\qquad B^{1/4}
\quad\mb{[GeV]}$  &\footnotesize$\qquad m_\chi \quad\mb{[GeV]}$   \\[1ex] 
\hline
$\qquad$2.0 &$\qquad$ 0.276 &$\qquad$ 1.502 \\[0.2mm]
$\qquad$2.1 &$\qquad$ 0.270 &$\qquad$ 1.409 \\
$\qquad$2.2 &$\qquad$ 0.264 &$\qquad$ 1.356 \\
$\qquad$2.3 &$\qquad$ 0.258 &$\qquad$ 1.337 \\
\hline
\end{tabular}
\end{center}
\begin{center}
  \parbox{0.9\textwidth}{ \refstepcounter{table} \footnotesize Table
    \thetable:{\label{tab1} UV cutoff dependence of the bag constant
      $B$ and glueball mass $m_\chi$. See text for details.}}
\end{center}
\end{figure}
The evolved minimum $\chi_0$ of the potential is shifted towards
smaller values for higher UV cutoffs. Due to the longer evolution with
increasing UV cutoff the absolute depth of the potential is also
increasing. Nevertheless the change for the bag constant is small (cf.
Tab.~\ref{tab1}) because, for its calculation via
Eq.~(\ref{bagkonst}), only the potential difference enters.  In
Tab.~\ref{tab1} the cutoff dependence of the glueball mass $m_\chi$ is
also listed. The decrease of the glueball mass with increasing UV
cutoff is more sensitive.

On the other hand if we keep the UV cutoff $\Lambda$ fixed and vary
the coupling $h$ or the minimum $\chi_0$ at the initial UV scale we
obtain in both cases a rather weak depencence of $B$ and $m_\chi$.
The glueball mass and the bag constant decrease with decreasing $h$.
The variation of $\chi_0$ is anyway restricted to a small interval
(within 7 \% of $\chi_0$) in order to avoid a pole in the threshold
function already mentioned above. Within this interval the glueball
mass decreases within 2\% while the bag constant increases within 2 \%
for increasing $\chi_0$.

\subsection{Finite-temperature evolution}

Within the Matsubara formalism, the finite-temperature version of the
self-consis\-tent flow Eq.~(\ref{flowg0}) is given by the following
expression~\cite{bohr,scha}

\ba 
\partial_t V_k
(\chi) &=& \frac{2\;\; \Gamma(3/2)}{ {(4\pi)^{(d-1)/2}} \Gamma(d/2)}
\frac{k^{d-1}}{d} T \!\!  \sum\limits_{n=-\infty}^\infty
\frac{1}{(1+\tilde{\omega}^2_n + \tilde{V}''_k)^{3/2}} \ .  
\label{FTflow}\ea

The non-integer powers in the threshold functions are typical for the
finite-temperature version within this approach and are governed by
the choice of the smearing functions $f_k$.  At finite temperature we
again solve, analogous to the zero-temperature case, the flow equation
(\ref{FTflow}) numerically by discretizing the $\chi$ field. We use
the same zero-temperature initial conditions at the UV scale $\Lambda
= 2.0$ GeV. Note that this choice restricts the finite-temperature
predictions of our results for higher temperatures.  The predictive
power of the finite-temperature extrapolation is basically determined
by the structure of the threshold functions in the flow equations and
the value of the UV cut-off. A detailed discussion can be found in
Ref.~\cite{scha}.  With $\Lambda = 2.0$ GeV we can ignore the
temperature dependence of the initial values at the UV scale up to
temperatures around $T \sim 250$ MeV.

\begin{figure}[!ht]
  \centerline{\hbox{
      \psfig{file=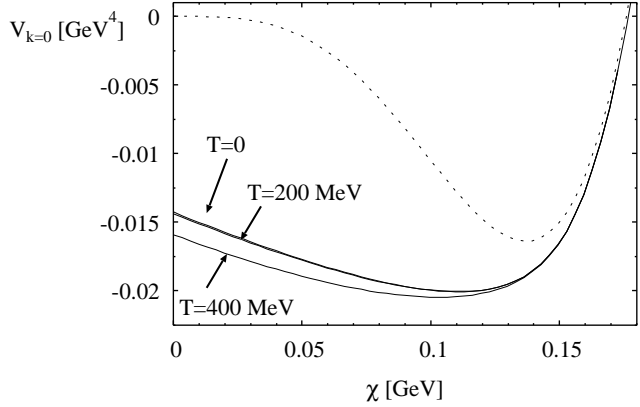, width=8.4cm,angle=0} }}
\caption{\label{figTpot}The potential $V_{k=0}$ (h$^{1/4}$
  = 0.51 GeV, $\Lambda = 2.0$ GeV , $\chi_0 = 0.137$ GeV ) versus the
  $\chi$ field for different temperatures ($\Delta T = 200$ MeV, see
  text for details).  }
\end{figure}

In Fig.~\ref{figTpot} we show the temperature behavior of the
potential in the IR. The dashed curve corresponds to the initial bare
potential at $k=\Lambda$. After the $k$ evolution we obtain the solid
curve, labeled by $T=0$ MeV, corresponding to the zero temperature IR
potential. We repeat this procedure in steps of $200$ MeV.  Below
temperatures of the order of $200$ MeV no significant changes in the
potential shape are observed.

\begin{figure}[!ht]
  \centerline{\hbox{
      \psfig{file=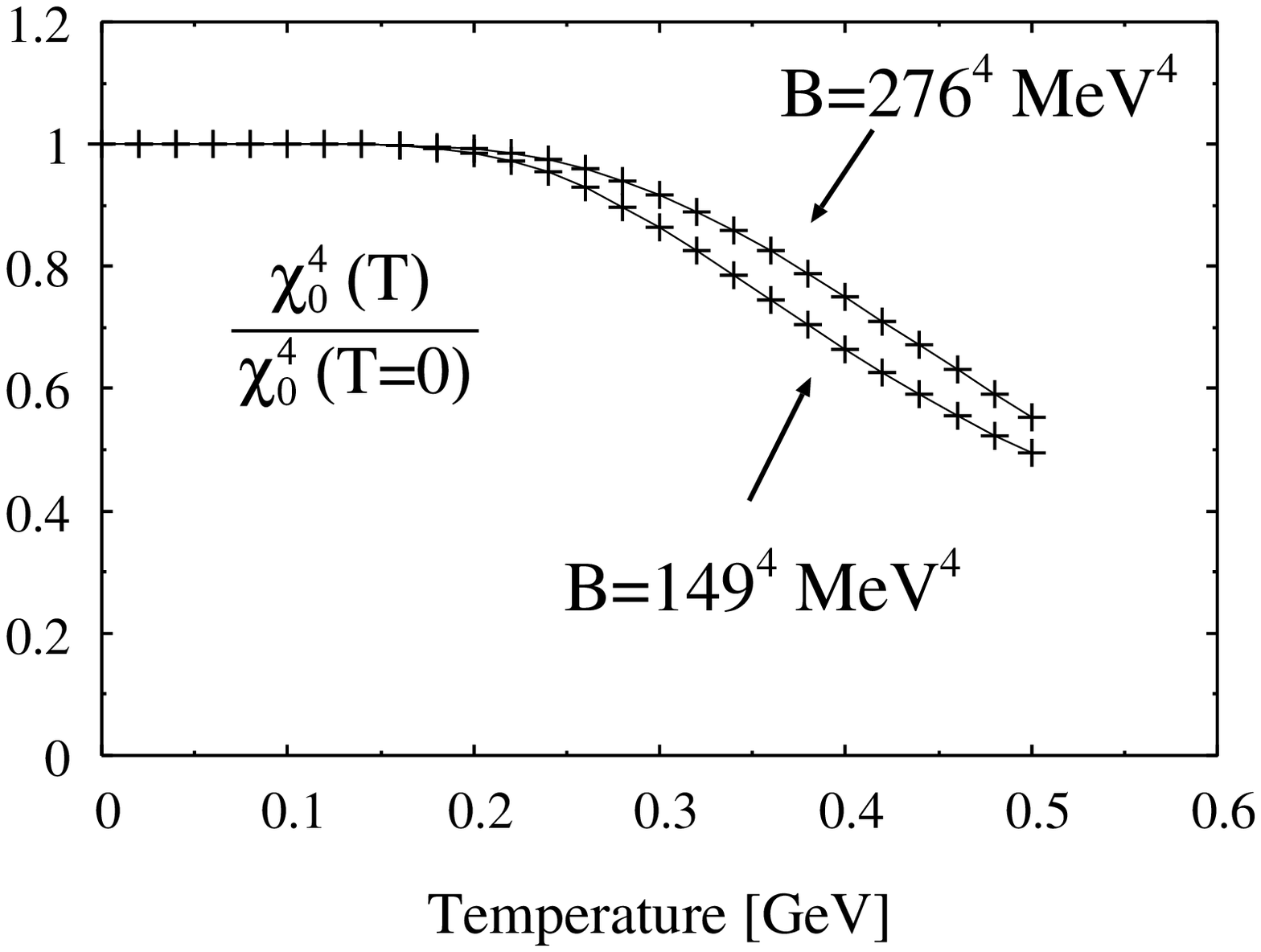,width=7cm,angle=0} }\hfill
    \psfig{file=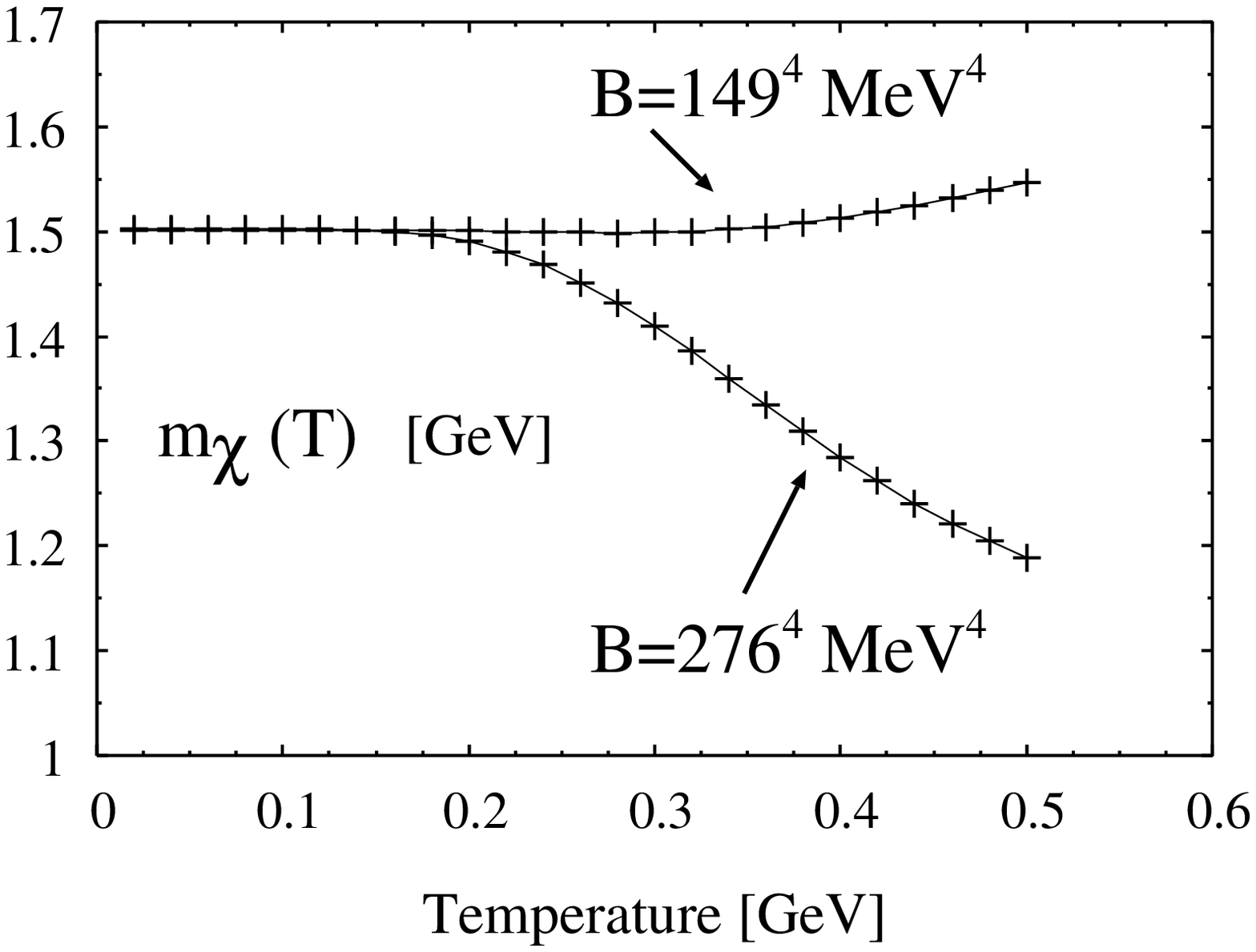,width=7cm,angle=0} }
\caption{\label{figTchi0}  
  The temperature-dependent normalized minimum $\chi_0^4 (T)/\chi_0^4
  (T=0)$ of the potential (left panel) and the glueball mass $m_\chi$
  (right panel) for two different bag constants (see text for
  details). }
\end{figure}
 
In analogy to the restoration of spontaneously broken chiral symmetry
at high temperature, we regard the temperature-dependent minimum of
the dilaton potential as an order parameter for the deconfining phase
transition under the assumption of factorization (see \cite{camp}).
Deconfinement would thus be signalled by a transition from a finite
VEV $\chi_0 \neq 0$ to a vanishing one $\chi_0 =0$. 

We observe a change of the normalized 'order parameter'
$\chi_0^4(T)/\chi_0^4(0) $ with temperature as shown in the left panel
of Fig.~\ref{figTchi0} for two different values of the bag constant
$B$ but equal glueball mass $m_\chi = 1.5$ GeV in the IR.  For both
bag constants the condensate stays almost constant up to temperatures
of the order $200$ MeV which is beyond the chiral phase transition
temperature for two quark flavors~\cite{kars}.  For the smaller bag
constant $B^{1/4} = 0.149$ GeV, which we obtain for $h^{1/4} = 0.34$
GeV and $\chi_0 = 0.068$ GeV at the UV scale $\Lambda = 2.0$ GeV, the
condensate starts to decrease earlier.  Thus, we can qualitatively
verify the dependence of the ``critical temperature'' on the bag
constant. In contrast to Ref.~\cite{soll} the onset of the shift in
the condensate is seen earlier at $T\sim 190$ MeV. It then decreases
almost linearly in both cases.  The negative slope for temperatures
beyond $300$ MeV does not depend on the chosen value for the bag
constant. At these temperatures we lose, however, predictive power due
to the omission of the temperature dependence of the initial values at
the UV scale (see also Ref.~\cite{scha}).  In the right panel the
temperature-dependent glueball mass $m_\chi$ is displayed for two
different bag constants.  The mass as function of temperature is
independent on the bag constant below temperatures of $180$ MeV. For
the larger bag constant $B^{1/4}=0.276$ GeV it then decreases but
increases again for very large temperatures above $800$ MeV. For the
smaller bag constant $B^{1/4}=0.149$ GeV the mass grows almost
linearly with temperature around $T\sim350$ MeV.  At very high
temperatures such a behavior is expected since perturbation theory is
applicable (cf.~\cite{cart}).  It turns out that this behavior is
almost independent of the initial $\Lambda,$ at least for temperatures
below $250$ MeV.

If we vary the cut-off $\Lambda$ between 1.5 and 2 GeV, while tuning
the coupling $h$ and $\chi_0$ to keep the glueball mass and the bag
constant fixed, we do not observe any significant changes below
temperatures of $200$ MeV.

Another more general way to find a possible RG fixed point solution
from a continuous flow equation starts with the rescaled fixed point
flow equation~\cite{morr}. For this purpose we introduce a
dimensionless potential $U( \bchi) = (4\pi)^2 k^{-4} V_k(\chi)$ and
field $\bchi = (4\pi) k^{-1} \chi$ in Eq.~(\ref{flowg0}) and set
$\partial_t U =0$ which defines the fixed point. In the LPA and $d=4$
this yields the flow equation

\ba
\label{fp} 
4 U(\bchi) - \bchi U' (\bchi ) &=& 
\frac{1}{2} \frac{1}{(1+U'' (\bchi))}\ .  
\ea

\begin{figure}[!ht]
  \centerline{\hbox{
      \psfig{file=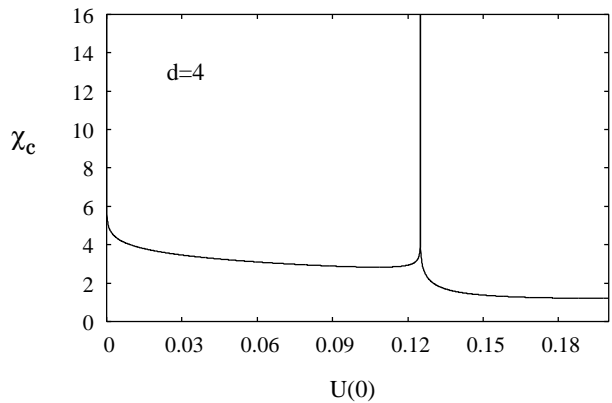,width=8cm,angle=0}}}
\caption{\label{figfp}
  The singularity $\chi_c$ as function of $U(0)$.}
\end{figure}

This equation has a continuum of solutions which depends on the
initial boundary conditions.  For the solution we have to specify two
boundary conditions. One is fixed by the necessary equation $U'(0) =
0$ at the critical point. If we choose some value for $U(0)$ we can
now numerically integrate Eq.~(\ref{fp}) out to positive fields
$\chi$. We almost always find a singularity at some critical field
$\chi = \chi_c$ where the numerics breaks down.  The result is
depicted in Fig.~\ref{figfp} where we find two exceptions.  The first
peak at $U(0)=0$ corresponds to a singular solution where $U''(0)$
diverges. Only for the second peak at $U(0) = 1/8$ the potential seems
to exist for all fields $\chi$. At this point the second derivative
vanishes and the potential itself is non-singular and constant
(trivial) for all fields $\chi$.

The peaks are generated by the threshold functions in Eq.~(\ref{fp})
and therefore depend on the choice of the smearing function. The
second solution is exactly at $U(0)=1/8$ for a smearing function
$f_k^{(m)}$ of the first kind $m=1$~\cite{bohr}.  Higher smearing
functions will influence the overall factor of the flow equations
moving the second peak towards zero like $2/(m+1)!$ \cite{bona}. We
conclude that no further non-trivial non-singular fixed point arises
within the LPA. A similar analysis for three space-time dimensions is
omited here since the extrapolation and implementation of the anomaly
is far from trivial.

\section{The truncated dilaton potential}
\label{truncg} 

In the previous section we have solved the non-truncated flow equation
(\ref{flowg0}) for the full dilaton potential in the LPA by
discretizing the scalar field $\chi$ on a grid.  Alternatively, one
can expand the potential in a Taylor series around the non-vanishing
local minimum $\chi_0$ (or any other field value) and derive flow
equations for the (in principle infinitely many) expansion
coefficients~\cite{aoki}. This allows for a direct study of the flow
of the coupling constants. We will investigate this procedure in this
section. Due to the termination of the potential series at a finite
power we expect new truncation effects. At any given finite order of
truncation there are always higher-order operators which do not evolve
and could thus influence the accuracy of the solutions.  In the limit
of infinitely many expansion coefficients we should, of course,
reproduce all results from the previous section.

The non-invariant Taylor expansion around the scale-dependent minimum
$\chi_0$ (broken phase) up to a specific order $M$ is given by

\ba 
V_k (\chi) &=& c_0 (k) + \sum\limits_{n=2}^M \frac{c_n
  (k)}{n!}  (\chi - \chi_0 (k))^n\ .  
\ea 

In the LPA the expansion coefficients, $c_n$, defined at the minimum
$\chi_0$, correspond to the $n$-point proper vertices evaluated at
zero momenta~\cite{polo}.  Substituting the potential expansion on
both sides of Eq.~(\ref{flowg0}) we can deduce a set of
coupled flow equations for the first several couplings $c_n,\ 
n=0,2,3,\ldots $.  \ba\label{truncflow}
\dot{c}_0 & = & \frac{1}{2(4\pi)^2}\frac{k^4}{1+ c_2/k^2}\nonumber \\
\dot{c}_1 & = & c_2\dot{\chi}_0 + \dot{c}_0\left[-\f{G_3}{k^2} \right]
= 0 \\
\dot{c}_2 & = & c_3\dot{\chi}_0 + \dot{c}_0\left[
  \frac{2G_3^2}{k^4} -  \frac{G_4}{k^2}\right] \nonumber\\
\dot{c}_3 & = & c_4 \dot{\chi}_0 + \dot{c}_0\left[ -6\frac{G_3^3}{k^6}
  + 6\frac{G_3 G_4}{k^4} - \frac{G_5}{k^2}
\right] \nonumber\\
\dot{c}_4 & = & c_5 \dot{\chi}_0 + \dot{c}_0\left[ 9\frac{G_3^4}{k^8}
  - 36\frac{G_3^2 G_4}{k^6} +8 \frac{G_3 G_5}{k^4}
  +6\frac{G_4^2}{k^4} -\frac{G_6}{k^2} \right] \nonumber\\
\dot{c}_5 & = & c_6 \dot{\chi}_0 + \dot{c}_0\left[
  -120\frac{G_3^5}{k^{10}} +240\frac{G_3^3 G_4}{k^8} -60\frac{G_3^2
    G_5}{k^6}-90\frac{G_3 G_4^2}{k^6} + \right. \nonumber\\
&\vdots& \qquad\qquad\qquad\qquad\qquad\qquad\qquad\left. + 10
  \frac{G_3 G_6}{k^4}+20\frac{G_4 G_5}{k^4} -\frac{G_7}{k^2} \right]
\nonumber \ea

\noindent
with the definition $G_n \equiv \frac{\D c_n}{\D (1+c_2/k^2)}$.  The
corresponding $\beta_n$-functions, $\beta_n \equiv k\frac{\D \partial
  c_n}{\D \partial k}=\partial_t c_n \equiv \dot{c}_n$, depend on
$\beta_n = \beta_n (c_0,c_1,\ldots,c_{n+2})$ for all $n$.  Due to the
absence of a reflection symmetry ($Z_2$ symmetry) we realize that all
odd and even vertices contribute to the $\beta_n$-functions.  The
expression for $G_n$ can be interpreted as a generalized $n$-point
vertex multiplied by a rescaled bosonic propagator with the squared
dilaton self-energy, $c_2$, given at vanishing external momenta
(cf.~\cite{papp}). This interpretation exemplifies the one-loop
contributions to the $\beta_n$-functions in a transparent and obvious
scheme.  For instance, the first few contributions in the brackets can
be visualized diagrammatically as
\begin{figure}[!ht] 
  \centerline{\hbox{ \psfig{file=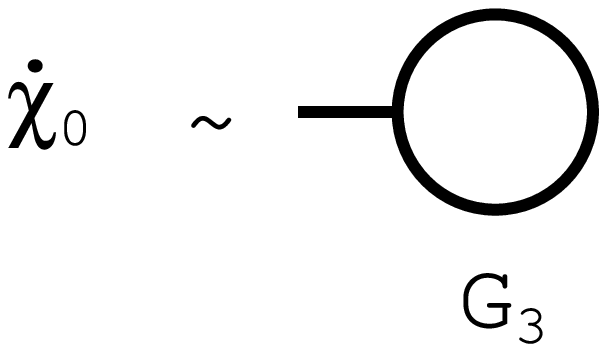,height=2.2cm,angle=0} }}
  \centerline{\hbox{ \psfig{file=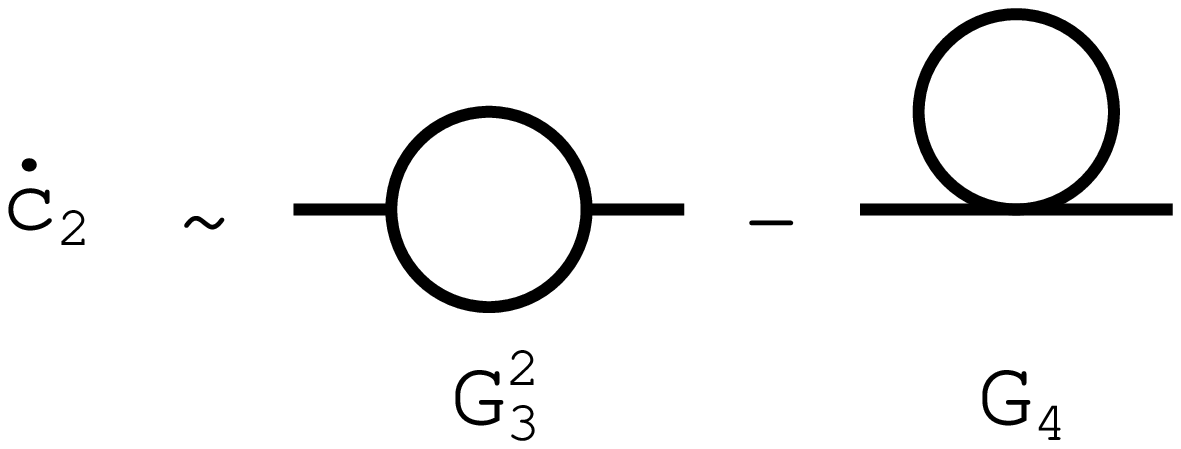,height=2.5cm,angle=0} }}
  \centerline{\hbox{ \psfig{file=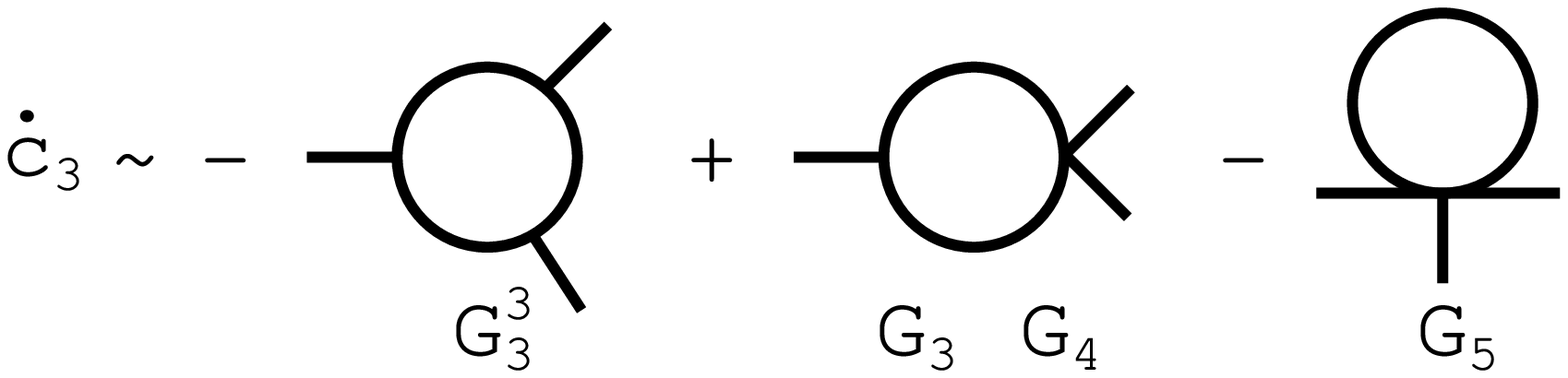,height=2.5cm,angle=0} }}
  \centerline{\hbox{ \psfig{file=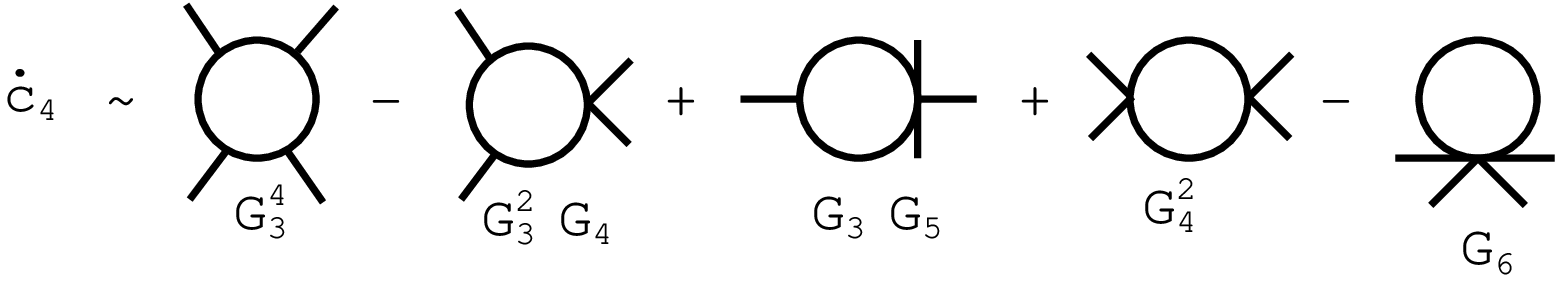,height=2.5cm,angle=0} }}
\caption{\label{figc2flow} Diagrammatical contributions to the first several
  $\beta_n$-functions. }
\end{figure}

\begin{figure}[!ht]
  \centerline{\hbox{ \psfig{file=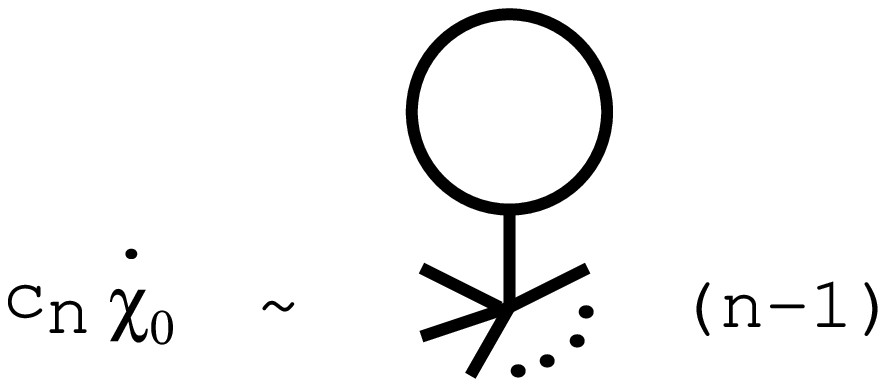,height=2.5cm,angle=0} }}
\caption{\label{figcnchi0} The term $c_n \dot{\chi}_0$ which has $(n -1)$ legs. }
\end{figure}

\noindent
One recognizes the one-loop structure.  In the diagrams we have
always omitted the first term on the $rhs$ of the $\beta_n$-functions
which is generically depicted in Fig.~\ref{figcnchi0}.
\\
     
The coefficient $c_1$ defines the minimum $\chi_0$ of the potential
and always vanishes: \be \left.V_k'\right|_{\chi_0} = c_1
\stackrel{!}{=} 0\ .  \ee

\begin{figure}[!ht]
  \centerline{\hbox{ 
      \psfig{file=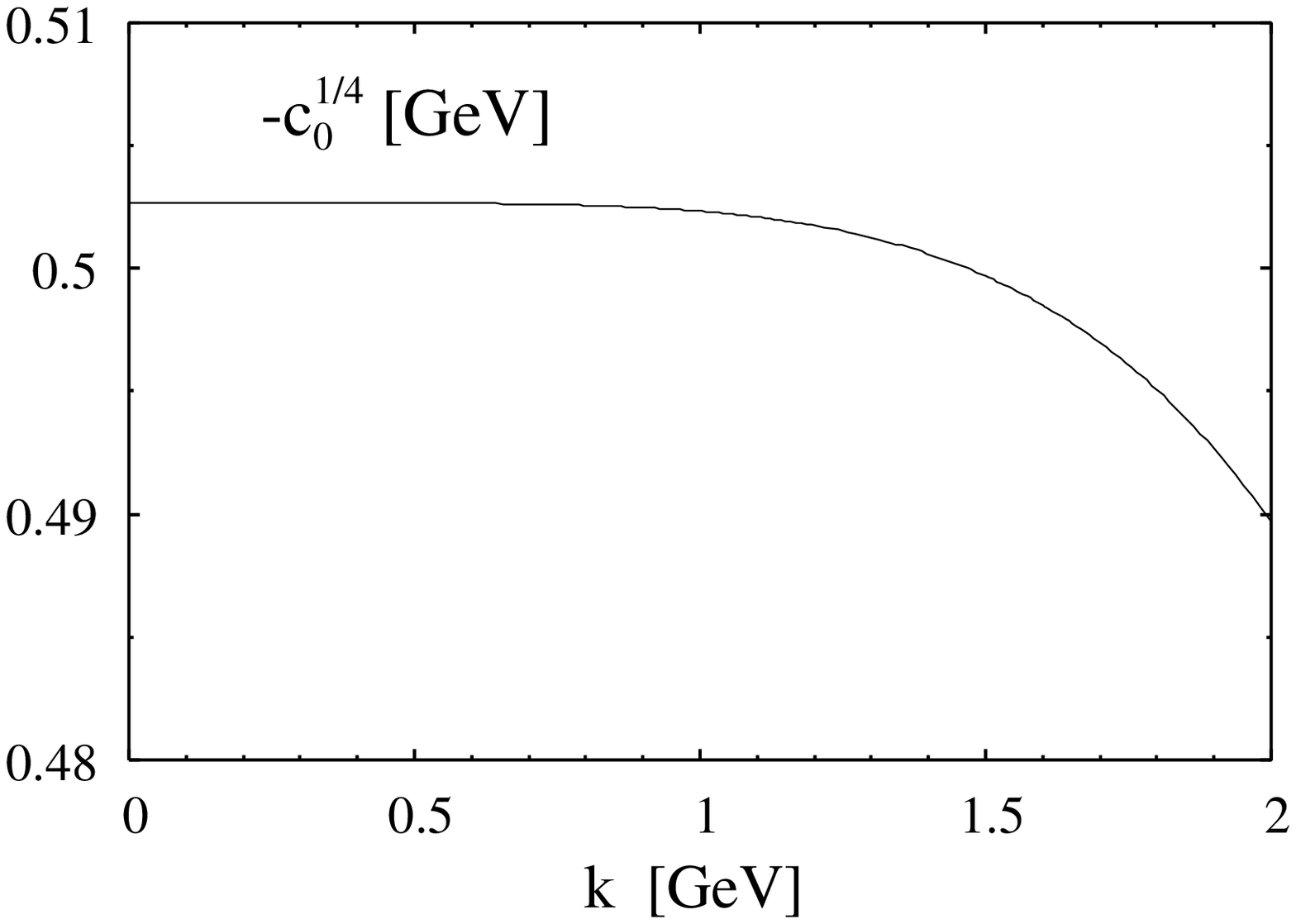,width=7.cm,angle=0}
      \psfig{file=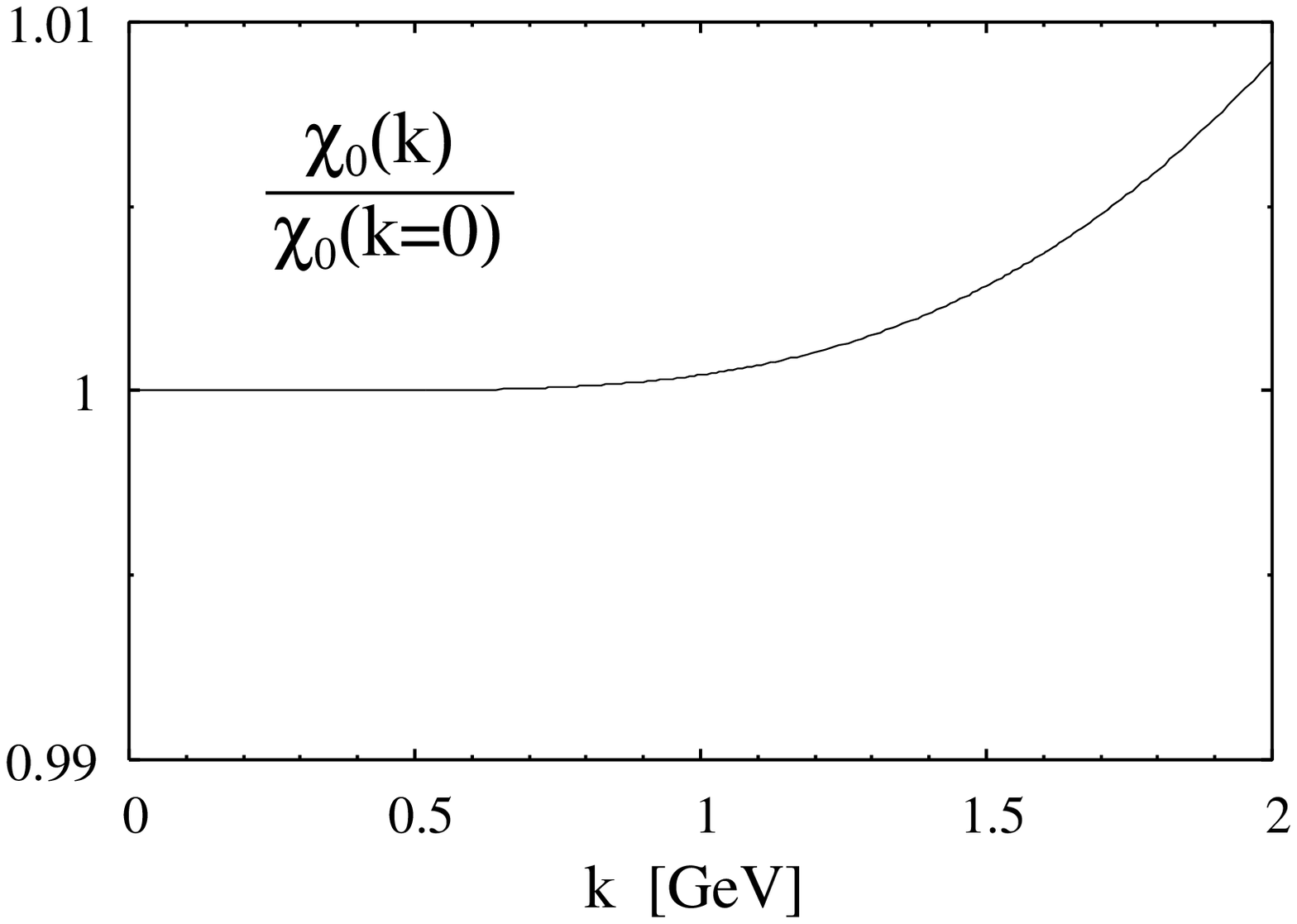,width=7.cm,angle=0} }}
  \centerline{\hbox{ 
      \psfig{file=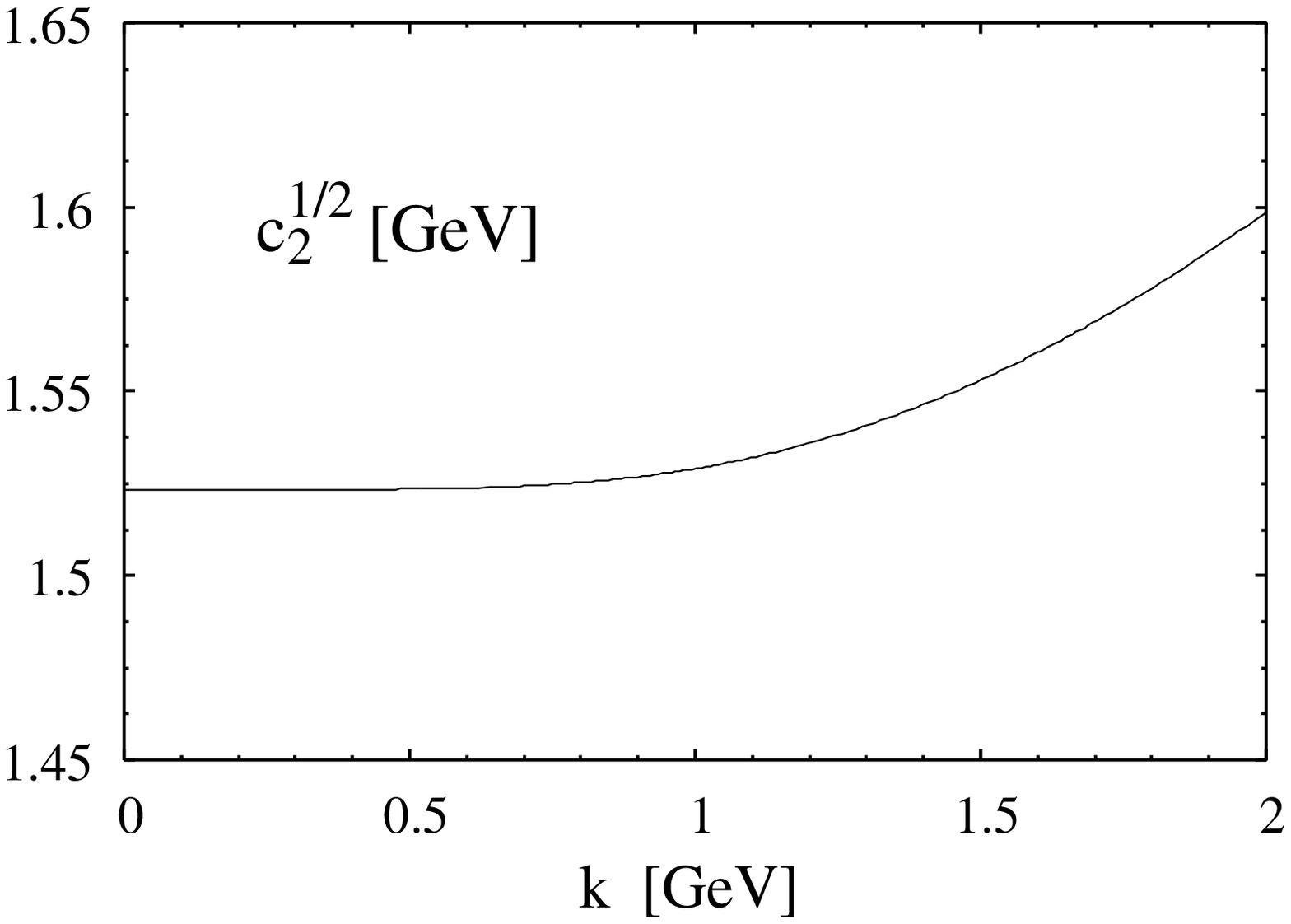,width=7.cm,angle=0}
      \psfig{file=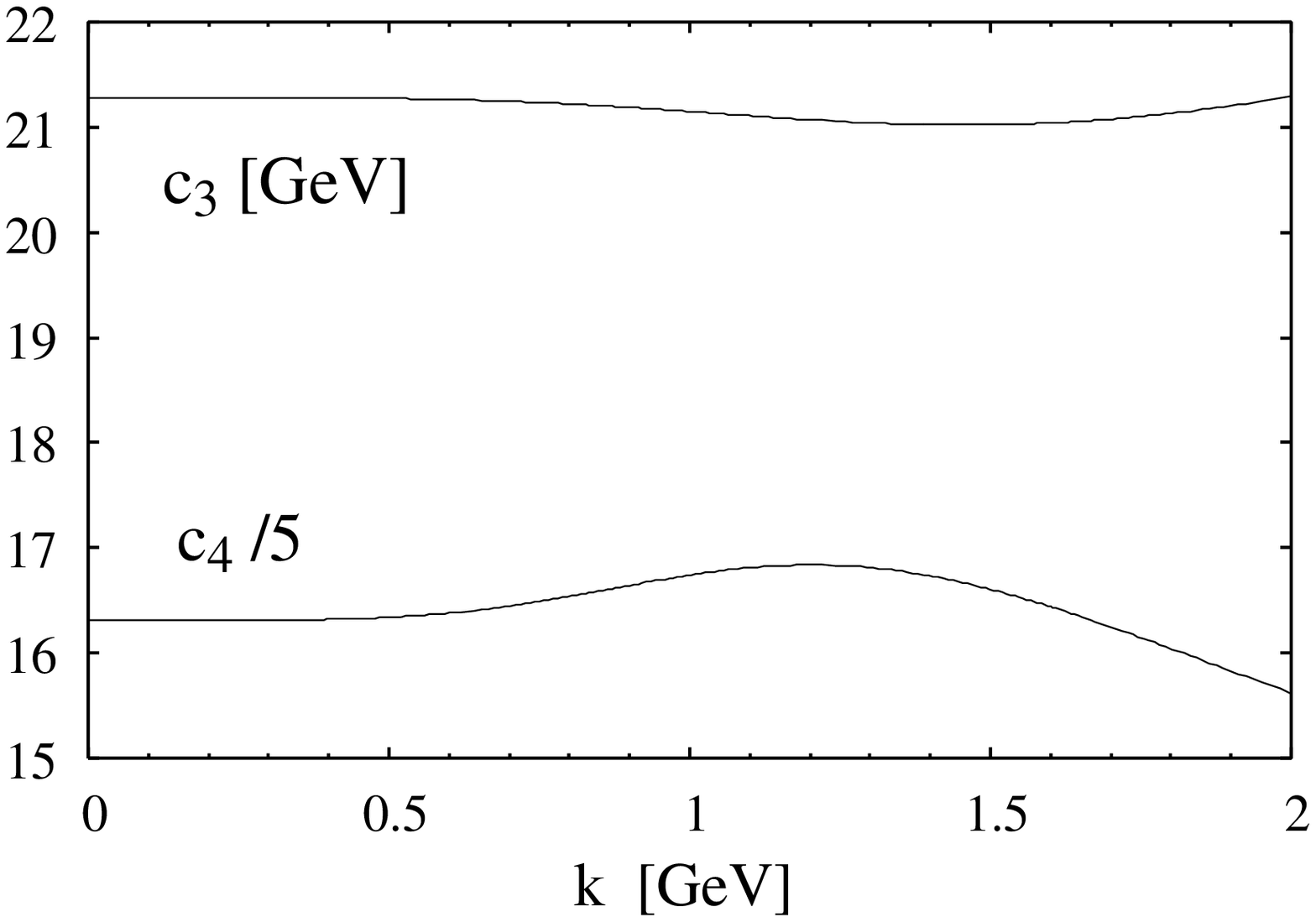,width=7.cm,angle=0} }}
\caption{\label{figtrunc} The scale evolution for the coefficients
  $-c_0^{1/4}$, $m_\chi = c_2^{1/2}$, $c_3$, $c_4$ and the minimum
  $\chi_0$. }
\end{figure}

We have solved the coupled flow equations (\ref{truncflow})
numerically with a 5th-order Cash-Karp Runge-Kutta method for
different truncation orders $M$ in order to investigate higher
coupling effects in detail.  As an example, in Fig.~\ref{figtrunc},
the $k$-evolution of the first coefficients $-c_0^{1/4}$, $c_3$,
$c_4$, the minimum $\chi_0$ and the dilaton mass $m_\chi = \sqrt{c_2}$
are shown for $M=5$. One observes that the $k$-evolution stops for all
dimensionful quantities around the scale $k \approx 400$ MeV. All
coefficients converge to finite IR values.

\begin{figure}[!ht]
  \centerline{\hbox{
      \psfig{file=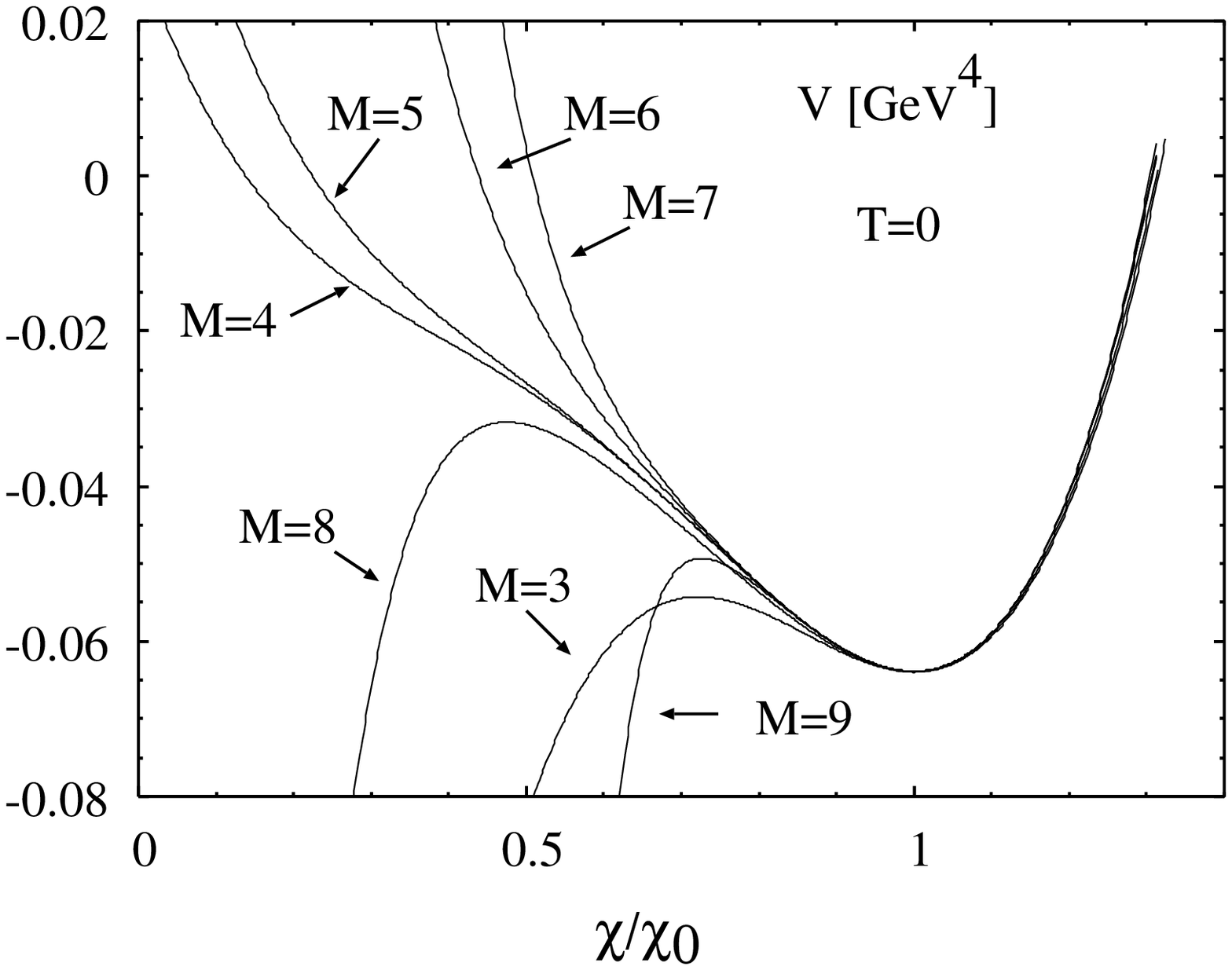,width=9.2cm,angle=0} }}
\caption{\label{figtruncpot} A comparison of the zero-temperature truncated
  potential calculation for different expansion order $M$,
  demonstrating the influence of higher operators to the potential.
  Each curve is labeled by the corresponding expansion order $M$.}
\end{figure}

In order to estimate the truncation effects we compare the truncated
potential calculation for different truncation orders $M$.  It turns
out that in the truncated potential calculation we cannot choose the
same initial values as for the full one in the previous sections.
With the values $\Lambda = 2$ GeV, $h^{1/4}=0.69$ GeV and $\chi_0 =
0.6$ GeV we can indeed perform the truncated potential calculation up
to the order $M=9$ where we have arbitrarily stopped our truncation
order. The results are presented in Fig.~\ref{figtruncpot}.  Each
curve depicts the evolved truncated potential in the IR up to a given
order $M$. Due to a slightly different evolution of the minimum
$\chi_0$ for different orders $M$ all potential curves are plotted
versus the $\chi$-field normalized in the IR.

\noindent
We do not observe an improved convergence of the truncated potential
to the full potential when increasing the order of the expansion.
Obviously, the truncation order $M=3$ is inappropriate.

\begin{figure}[!ht]
  \centerline{\hbox{
      \psfig{file=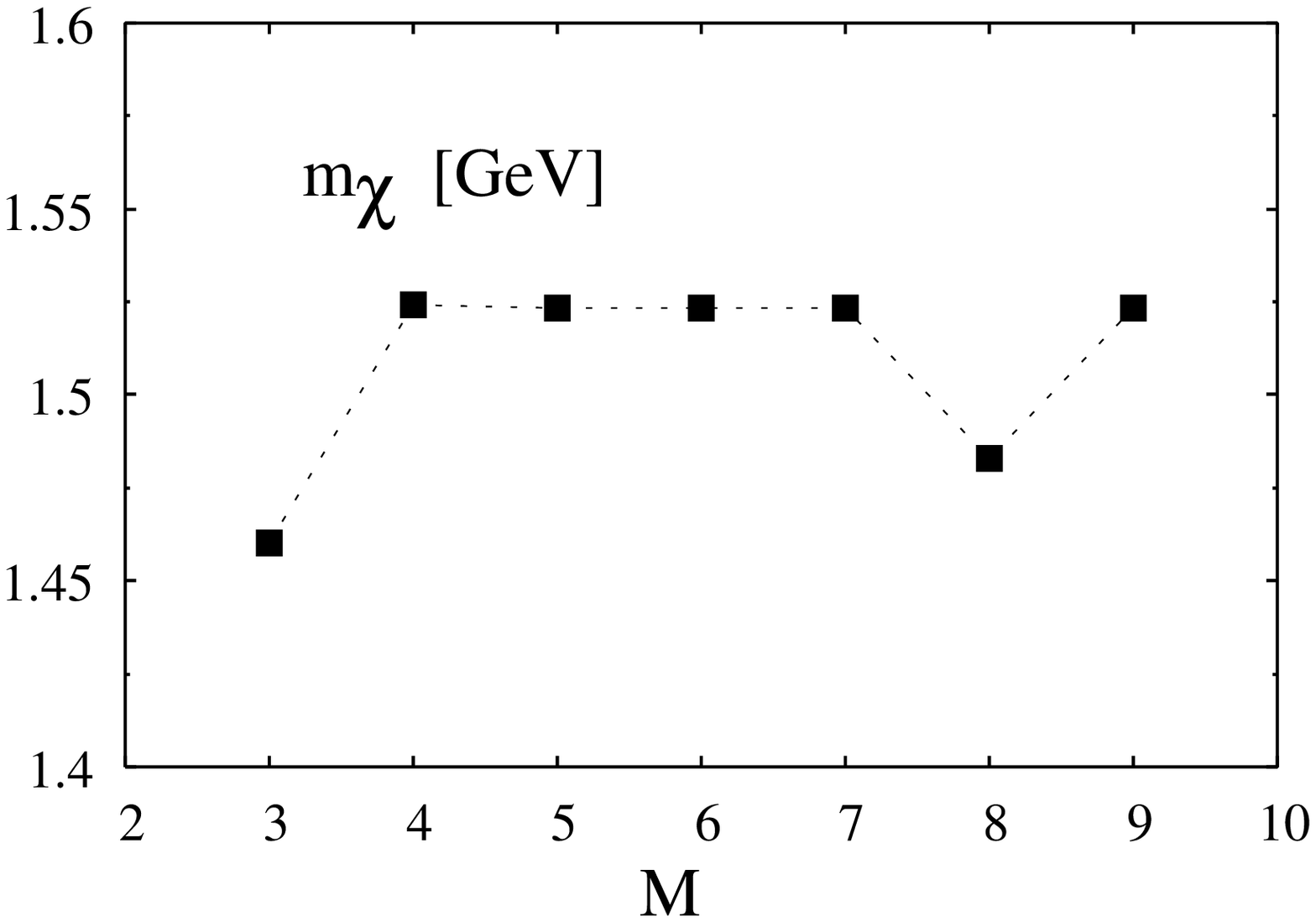,width=9cm,angle=0} 
      }}
\caption{\label{figmassM} The glueball mass $m_\chi$ as function of
  the order, $M$, of the polynomial expansion.}
\end{figure}

\noindent
In Fig.~\ref{figmassM} the evolved IR glueball mass, $m_\chi$, is
shown for different truncation orders $M$. All masses are obtained by
the same initial values thus displaying the influence of higher
operators on the mass evolution. We again do not see a convergence up
to the order $M=9$.

\begin{figure}[!ht] 
  \centerline{\hbox{ \psfig{file=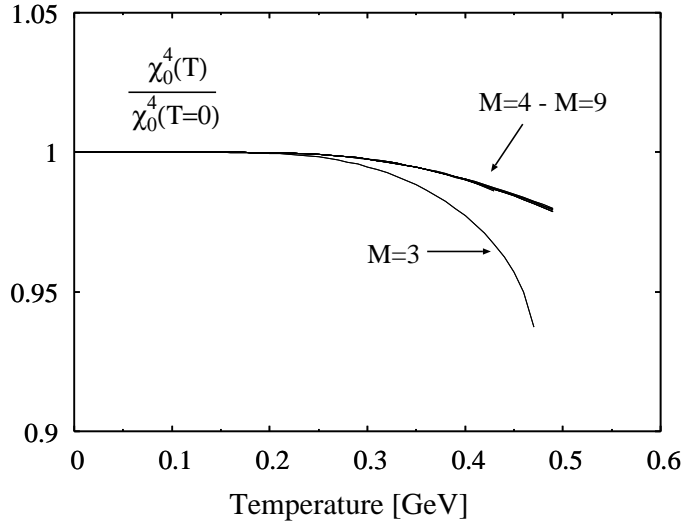,
        width=9cm,angle=0}
      }}
\caption{\label{figtruncchi} 
  The temperature-dependent normalized minimum $\chi_0^4 (T)/\chi_0^4
  (T=0)$ of the potential for different truncation orders $M$.
  ($\Lambda = 2$ GeV, $h^{1/4} = 0.69$ GeV, $\chi_0 (\Lambda)
  =0.6$ GeV).}
\end{figure}

The finite-temperature generalization of the expansion procedure is
straightforward. In Fig.~\ref{figtruncchi} we show the normalized
minimum of the truncated potential at finite temperature for different
values of $M$.  

For all orders the minimum stays constant up to temperatures around
$200$ MeV. For higher temperatures we see truncation effects due to
the omission of higher operators in the corresponding beta-functions.
The rapid temperature decrease of the minimum for truncation order
$M=3$ could also be related to the smaller zero-temperature glueball
mass at this order.  Increasing the expansion order $M$ the minimum
becomes more stable.  Even when varying the UV cut-off scale from
$\Lambda =2$ GeV towards $\Lambda=1.5$ GeV while fixing the glueball
mass we do not see strong deviations (within 0.1 \%) in the normalized
minima around temperatures of $200$ MeV.

We therefore believe that we can trust the expansion of the dilaton
potential up to a finite order $M$ when we are interested in
temperatures below $200$ MeV. In a forthcoming publication \cite{wamb}
we are going to investigate the finite-temperature behavior of a
scaled linear sigma model and study the influence of mesons on this
result.

\section{Conclusions and Outlook}
\label{concl}

We have investigated the thermal behavior of the gluon condensate
within a self-consistent proper-time regularized Renormalization Group
approach. An effective realization of the QCD trace anomaly by a
scalar dilaton field is used, which yields a logarithmic potential
parameterized by two constants.  Due to the breakdown of the effective
model for vanishing dilaton field, a perturbative analysis is not
possible. Therefore, the order of the phase transition cannot be
determine perturbatively.  Using non-perturbative RG flow equations
these difficulties can be circumvented and the calculation of the
effective potential at the origin becomes possible.

We have considered two procedures to explore a possible phase
transition in the effective dilaton model. First the full logarithmic
potential is considered. The analysis is then compared to a truncated
potential calculation where truncation errors are observed and
discussed. The full potential analysis yields the potential for all
dilaton fields and not only around the minimum.  In contrast, the
truncated potential calculation applies only in the vicinity of the
global minimum.

At vanishing temperature, the model has been fixed to the glueball
mass and bag constant. Our finite temperature predictions for the full
potential are in good agreement with the results of Ref.~\cite{soll}
while the truncated potential calculation suffers from slow
convergence and artificial truncation effects. Yet, the finite
temperature results of the truncated potential calculation correspond
qualitatively to those of the full potential analysis. One advantage
of the truncated version is that a diagrammatical interpretation of
the beta functions in terms of $n$-point vertices becomes possible,
elucidating the underlying physics more clearly.  The gluon condensate
and the glueball mass remain unaltered up to temperatures of about 200
MeV where our prediction reaches its limit of validity due to the
omission of temperature-dependent initial boundary conditions in the
flow equations.

In contrast to~\cite{camp} we do not observe any phase transition
within the considered temperature range which makes the use of the
expectation value of the dilaton field as an order parameter for the
gluon deconfinement questionable.  On the other hand this result is
not astonishing, since the trace anomaly is not expected to vanish at
high temperature~\cite{leut}.
 
We have neglected the influence of mesonic degrees of freedom which
should anyhow play a minor role on the gluonic phase transition
\cite{soll,agas,cart}. The consideration of these degrees of freedom
in this RG framework is deferred to a forthcoming publication
\cite{wamb}.

\end{document}